\documentclass[12pt]{article}

\usepackage{enumitem} 

\usepackage{setspace} 

\usepackage{amssymb,amsmath,amsfonts,bbm}
\usepackage{graphicx}
\usepackage{mathrsfs}
\usepackage{calligra}
\usepackage{multirow}
\usepackage[OT4]{fontenc}

\numberwithin{equation}{section}

\newcommand{\PMB}[1]{
\leavevmode
\setbox0=\hbox{#1}%
\kern-0.02em\copy0\kern-\wd0
\kern0.02em\copy0\kern-\wd0
\kern-0.025em\raise0.0167em\box0
\kern-0.025em\raise0.0433em\box0}

\begin{document}

\begin{center}
{\Large {\tt Diagnostic tools for exploring differences in distributional properties between two samples: nonparametric approach}}\\

\vspace{1.5cm}
{\sf Bogdan \'Cmiel}\\
{Faculty of Applied Mathematics, AGH University of Science and Technology,\\ Al. Mickiewicza 30, 30-059 Cracov, Poland}\\
and\\
{\sf Teresa Ledwina}\\
{Institute of Mathematics, Polish Academy of Sciences,\\ ul. Kopernika 18, 51-617 Wroc{\l}aw, Poland}
\end{center}

\vspace{1.5cm}

\noindent
{\bf Summary}\\

\noindent
This paper reconsiders the problem of testing the equality of two unspecified continuous distributions. The framework, which we propose, allows for readable and insightful data visualisation and helps to understand and quantify how two groups of data differ. We consider a useful weighted rank empirical process on (0,1) and utilise a grid-based approach, based on diadic partitions of (0,1), to discretize the continuous process and construct local simultaneous acceptance regions. These regions help to identify statistically significant deviations from the null model. In addition, the form of the process and its dicretization lead to a highly interpretable visualisation of distributional differences. We also introduce a new two-sample test, explicitly related to the visualisation. Numerical studies show that the new test procedure performs very well. We illustrate the use and diagnostic capabilities of our approach by an application to a known set of neuroscience data.\\

\vspace{0.5cm}


\noindent
{\sf Email adresses:} cmielbog@gmail.com (Bogdan \'Cmiel), ledwina@impan.pl (Teresa Ledwina)\\

\newpage
\noindent
{\bf Content}\\

\noindent
1. Introduction \dotfill 2\\
2. Preliminaries: notation, contrast comparison curve and related weighted rank process \dotfill 3\\
3. Inspection and evaluation of the difference between $F_m$ and $G_n$ \dotfill 4

3.1. Discretization and B-plots \dotfill 5

3.2. Local comparisons \dotfill 5

3.3. Global test $\sf Max_{D(N)}$ pertinent to B-plots \dotfill 7

3.4. Real data examples \dotfill 7

\noindent
4. Simulated powers of $\sf Max_{D(N)}$ and some of its competitors \dotfill 11

4.1. Compared tests \dotfill 11

4.2. Alternative models and empirical powers \dotfill 12\\
\noindent
5. Discussion \dotfill 15\\
\noindent
Appendix A: Contrasting two-rank processes under $F=G$ \dotfill 16\\
\noindent
Appendix B: Proofs \dotfill 20\\
\noindent
References \dotfill 22\\

\noindent
{\bf 1. Introduction}\\

Nonparametric two-sample procedures are useful and very popular in several areas of application, including biomedical and insurance data. In our contribution, we focus on univariate data from continuous population cumulative distribution functions $F$ and $G$, say. In spite of their own recognised importance,  univariate two-sample statistics are building blocks in change-point detection and in some multivariate comparison procedures. The purpose of the presented study is to answer whether or not the two populations differ and, if so, to understand how they differ.

In the univariate case, many tests have been constructed to decide whether the two distributions $F$ and $G$ are identical or different. Thas (2010) discusses several of them. Simultaneously, graphical methods have been developed both for informal and formal inference about differences between pertinent two empirical distributions. The work of Fisher (1983) nicely summarises important early developments. Many contemporary papers still deal with elaborating procedures that allow us to understand distributional differences between two groups of measurements and to visualise them. For an illustration, see de Jong et al. (1994), Wilcox (1995), Duong (2013), Rousselet et al. (2017), Goldman $\&$ Kaplan (2018), Brown $\&$ Zhang (2024), Ledwina $\&$ Zagda\'nski (2024), Konstantinou et al. (2024) and references therein.

This paper introduces a novel nonparametric method for visualising and statistically evaluating the differences in distributional properties between two samples. In our contribution, we follow the approach introduced in Ledwina $\&$ Wy{\l}upek (2012a,b) and developed in  Ducharme $\&$  Ledwina (2024) and Ledwina $\&$ Zagda\'nski (2024). The focus is on differences in population distributions on a quantile scale. In contrast to may contributions dealing with immediate comparison of quantile functions or selected quantiles, in our interdistributional comparisons we relay on relative distributions. This provides a popular setting in many applications and results in stable, accurate and insightful methods of assessing how the distributions $F$ and $G$ differ. Our first step is to provide data visualisation to guide further analysis. For this purpose, the contrast comparison curve, introduced in Ledwina $\&$ Zagda\'nski (2024), is used. Its natural estimate is a weighted rank empirical process, which is the basis of our procedures. The first kind of procedure, we discuss, are simultaneous local acceptance regions which allows us to assess significance of some local departures between two empirical cumulative distribution functions (cdf's, in short). Next, a new global test related to the comparison curve is introduced. More precisely, this article is organised as follows.

First, by way of background, we present recently introduced notion of the contrast comparison curve, related rank empirical process, and comment on advantages of such approach over much more popular in applied statistics use of classical empirical ROC process. Second, also as a background, we provide adequate variants of simultaneous local acceptance regions and weighted max-type test for assessing equality of two distributions. Third, we show an application of our approach to real data from neuroscience and compare the results with the results of the analysis presented by Wilcox and Rousselet (2023).
We conclude with extensive simulation comparison of our max-type statistic with recent ROC-based solution proposed by Tang et al. (2021) and the classical integral Anderson-Darling statistic.  The results of the real data analysis and the simulation study are very encouraging. 
Appendix A provides a comparison of the stability of our setting with a more traditional one based on a classical empirical ROC process. The proofs are collected in Appendix B. \\

\noindent
{\bf 2. Preliminaries: notation, contrast comparison curve and related weighted rank process}\\

We consider two independent samples $X_1,...,X_m$ and $Y_1,...,Y_n$ obeying continuous cdf's $F$ and $G$, respectively. Set $N=m+n,\;\lambda_N=m/N$ and $H(z)=\lambda_NF(z) + (1-\lambda_N)G(z),\;z \in \mathbb{R}$. 
Ledwina $\&$ Zagda\'nski (2024) have introduced the curve
$$
\text{CCC}(p)= \frac{F(H^{-1}(p))-G(H^{-1}(p))}{\sqrt{p(1-p)}}=
\frac{p-G(H^{-1}(p))}{\lambda_N \sqrt{p(1-p)}},\;\;\;p \in (0,1),
\eqno(1)
$$
and called it the {\it contrast comparison curve}. Its origins and properties were extensively discussed therein. We recall some of the properties in the course of real data analysis in Section 3.4. Several graphs of CCC curves for standard pairs of $F$ and $G$ are shown in Section 4.2.

To define the natural estimate $\widehat{\text{CCC}}$ of the CCC set $F_m$ and $G_n$ to be the right-continuous empirical cdf's based on the first and the second samples, respectively. For the pooled sample $(X_1,...,X_m,Y_1,...,Y_n)$, let $H_N=\lambda_NF_m + (1-\lambda_N)G_N$ denote the pertinent empirical cdf. Then
$$
\widehat{\text{CCC}}(p)=w(p)[F_m(H_N^{-1}(p))-G_n(H_N^{-1}(p))],\;\;\;w(p)=1/\sqrt{p(1-p)},\;\;\;p \in (0,1).
\eqno(2)
$$
Introduce also two processes related to (2). Namely, the unweighted variant
$$
{\widehat P}_N(p)=\eta_N [F_m(H_N^{-1}(p))-G_n(H_N^{-1}(p))],\;\;\;p \in [0,1],\;\;\;\eta_N=\sqrt{mn/N},
\eqno(3)
$$
and the weighted variant 
$$
{\sf P}_N(p)= \eta_N w(p) [F_m(H_N^{-1}(p))-G_n(H_N^{-1}(p))],\;\;\;p \in (0,1).
\eqno(4)
$$ 
The process ${\widehat P}_N(p)$ has been exploited by Behnen and Neuhaus in their developments on rank tests with estimated scores, cf. Behnen $\&$ Neuhaus (1983) for an illustration, and in some data-driven tests constructions; see Janic-Wr\'oblewska $\&$ Ledwina (2000) for an illustration. In turn, the process ${\sf P}_N(p)$ was exploited in Ledwina $\&$ Wy{\l}upek (2012a,b) and successive papers on some variants of one-sided testing problems. In the present paper, we shall also rely on this weighted process. 

It should be noted that the literature on comparing two distributions is strongly dominated by an ODC/ROC-based approach. The empirical ROC process $\widehat{R}_N(p)=\eta_N [G(F^{-1}(p))-G_n(F_m^{-1}(p))]$
plays a central role in this approach. It is known that, under $F \neq G$, the asymptotic variance function of $\widehat{R}_N(p)$ is unbounded in many cases of $F$ and $G$. This is in sharp contrast to the variance function of ${\widehat P}_N(p)$, which is always bounded. On the other hand, under $F=G$, both processes obey the same asymptotic distribution with variance $p(1-p)$. Related results are collected in Appendix A of Ledwina $\&$ Zagda\'nski (2024). Moreover, extensive simulations reported in Ledwina $\&$ Zagda\'nski (2024) clearly show that, under $F=G$, the finite sample distributions of $\widehat{R}_N(p)$ are much more variable than those of $\widehat{P}_N(p)$, when $p$ is close to 0 and 1. 
The phenomenon is especially strongly manifested when the sample sizes $m$ and $n$ are highly unbalanced, even when the sizes are considerably large. We provide some explanation of this instability in Appendix A of the present contribution. The above-mentioned evidence makes
$\widehat{P}_N(p)$ a much more stable base for the test construction of ${\mathbb H} : F=G$ versus ${\mathbb A} : F \neq G$ than the process $\eta_N [p-G_n(F_m^{-1}(p))]$ does.\\

\noindent
{\bf 3. Inspection and evaluation of difference between $F_m$ and $G_n$}\\

As mentioned in Section 1, our approach is based on developments elaborated in Ledwina $\&$ Wy{\l}upek (2012a,b), where some smooth components, related to projected Haar functions, were introduced and combined to define some data-driven Neyman tests and some max-type statistics for interdistributional comparisons of two samples in two different one-sided testing problems. However, it should be admitted that many other view points, discussed elsewhere,  can result in similar solutions. In the present paper, we follow the exposition proposed in Ledwina $\&$ Zagda\'nski (2024) and think in terms of stochastic process ${\sf P}_N(p)$ and its discretization. Moreover, simultaneous local acceptance regions, invented recently in Ducharme and Ledwina (2024) in the context of goodness-of-fit testing, are adjusted to the present two-sided nonparametric testing problem. 

Two basic features of our approach are: discretization of inspection points $p$'s and restriction of the range of $p$'s to some useful subinterval;  depending on underlying sample sizes $m$ and $n$ in our implementation. The question of discretization is popular nowadays. For some literature, see Section 3 of Ledwina $\&$ Zagda\'nski (2024) and Section 4.1 in \'Cmiel $\&$ Ledwina (2024). The idea of restriction of the range of inspection points in some weighted suprema goes back to the seminal paper by Borovkov $\&$ Sycheva (1968) and has been rediscovered and exploited in some application-orientated papers as well. 

In the following, we give some details and show an application of our tools to known sets of real data.\\

\noindent
{\sf 3.1. Discretization and B-plots}\\

When starting with CCC(p) and the continuous-time process ${\sf P}_N(p)$, then, in principle, any convenient discretization can be introduced. We use the discretization following the diadic partition of [0,1], as it is explicitly related to the initial ideas proposed in Ledwina $\&$ Wy{\l}upek (2012a,b), and has appeared to be useful in several applications. To introduce the needed details, we start with some notation related to the discretization.  Given a resolution level $s$, $s=0,1,...$, we set $d(s)=2^{s+1}-1$ and define the points 
$$
p_{s,j}=\frac{j}{2^{s+1}}=\frac{j}{d(s)+1},\;\;\;j=1,...,d(s).
$$
For the given total sample size $N=m+n$, we shall restrict our attention to $s=S(N)$ and $D(N)=d(S(N))$.

Having selected the points in (0,1), in our inferential procedures, we shall rely on the values of the process ${\sf P}_N(p)$ over these points. Representing thus obtained values as bars results in the pertinent {\it bar plot} (B-plot, in short). Taking into account that, under $F=G$, each bar is asymptotically $N(0,1)$, B-plot provides the first insight if and where the difference between $F$ and $G$ seems to contradict their hypothetical equality. On the one hand, it should be admitted that the normal approximation of one-dimensional distributions of ${\sf P}_N(p)$, under $F=G$, is satisfactory, even for relatively small sample sizes, except $p$'s very close to 0 and 1. So, we can almost immediately infer which bars have an unlikely magnitude. However, the bars are correlated. Therefore, to obtain some reliable conclusions on a group of bars, this fact should be included in our argument. For this purpose, we shall use simultaneous local acceptance regions defined below.\\

\noindent
{\sf 3.2. Local comparisons}\\

Local comparisons serve us to understand: ``How do observations in specific parts of a distribution compare between groups'', as the question posed by Rousselet et al. (2017), p. 1740.
Though some ideas have been published relatively early in papers on applied statistics, cf. Campbell (1994) and Wilcox (1995), for example, it seems that a more wide interest in this aspect of two-sample comparison in the area of mathematical statistics has been noticed recently only. The latest papers by Brown and Zhang (2024), Ledwina and Zagda\'nski (2024), and Konstantinou et al. (2024) discuss some approaches and provide many related references. 

In our contribution, we follow the development in Ducharme $\&$ Ledwina (2024), adjusted to the two-sample case in Ledwina $\&$ Zagda\'nski (2024). The last mentioned implementation was tailored taking into account some discussions in the econometric literature. In general, our idea is to consider {\it simultaneous local acceptance regions} (local acceptance regions; in short) related to discretization. Our construction is as follows.

\begin{itemize}
\item Restrict your attention to successive deciles of $H_N(x)$. Hence, consider ten intervals $I_1=[0,0.1],\;I_2=(0.1,0.2],...,I_{10}=(0.9,1]$.
\item Next, given $I_k$, introduce
$$
L^{-}({\sf P}_N,I_k)=\min_{j \in I_k} {\sf P}_N(p_{S(N),j})\;\;\;\mbox{and}\;\;\;L^{+}({\sf P}_N,I_k)=\max_{j \in I_k} {\sf P}_N(p_{S(N),j})
$$
to denote the local minima and local maxima of the discretized process ${\sf P}_N(p)$.
\item Given $\alpha \in (0,1)$ and an interval $I_k$, under $F=G$, calculate the barriers ${l}^{-}(N,\alpha/2, I_k)$ and ${l}^{+}(N,\alpha/2, I_k)$ defined as follows
$$
P \Bigl(L^{-}({\sf P}_N,I_k) \geq  {l}^{-}(N,\alpha/2, I_k)\Bigr) \geq 1-\alpha/2
$$
and \hfill{(5)}
$$
P \Bigl(L^{+}({\sf P}_N,I_k) \leq  {l}^{+}(N,\alpha/2, I_k)\Bigr) \geq 1-\alpha/2.
$$ 
\end{itemize} 
Since the process ${\sf P}_N$ is distribution free under $F=G$, ordinary Monte Carlo suffices to calculate barriers ${l}^{-}(N,\alpha/2, I_k)$ and ${l}^{+}(N,\alpha/2, I_k)$. Moreover, (5) yields
$$
P \Bigl({l}^{-}(N,\alpha/2, I_k) \leq L^{-}({\sf P}_N,I_k),\; L^{+}({\sf P}_N,I_k) \leq {l}^{+}(N,\alpha/2, I_k)\Bigr) \geq 1-\alpha.
\eqno(6)
$$
Hence, $[{l}^{-}(N,\alpha/2, I_k),\; {l}^{+}(N,\alpha/2, I_k)]$ is simultaneous local acceptance region (for the bars in $I_k$) at the level $1-\alpha.$

Throughout, we consider $\alpha=0.05$. Note also that (5) defines two one-sided local acceptance regions for testing $F \geq G$ and $F \leq G$, respectively. In turn, (6) represents an adequate acceptance region in the case of verifying $F=G$.  In Section 3.4, we propose to represent these three simultaneous acceptance regions using shaded strips in the B-plot, applying light red, light blue, and white colours, respectively. \\

\noindent
{\sf 3.3. Global test ${\sf Max_{D(N)}}$ pertinent to B-plots}\\

For testing $F=G$ on the basis of B-plot pertinent to ${\sf P}_N(p)$, one of natural solutions is to consider 
$$
{\sf {Max}_{D(N)}} =
\max_{1 \leq j \leq D(N)} |{\sf P}_N(p_{S(N),j})| =
								\eta_N \times \max_{1 \leq j \leq D(N)}\frac{|F_m(H_N^{-1}(p_{S(N),j}))-G_n(H_N^{-1}(p_{S(N),j}))|}{\sqrt{p_{S(N),j}(1-p_{S(N),j})}}.
$$
This statistic is a counterpart of the one-sided statistic $M_{D(N)}$ introduced and studied in Ledwina $\&$ Wy{\l}upek (2012a, 2013). Note also that, in contrast to $M_{D(N)}$, we are not using the continuity correction in the present construction. ${\sf Max_{D(N)}}$ is similar in character to the well-known weighted two-sample Kolmogorov-Smirnov-type tests considered in Doksum $\&$ Sievers (1976) and Miller $\&$ Siegmund (1982), among others. Nevertheless, a closest analogue is Anderson's (1996) solution which includes both some discretization and related quantile hits. However, in contrast to Anderson's setting, we propose to consider much more dense inspection points $p_{S(N),j}$'s and allow their number to tend to infinity as $N \to \infty$.

The interpretation of ${\sf Max_{D(N)}}$ is simple and intuitive. It returns the maximized (reweighted) difference between $F_m$ and $G_n$ on selected quantiles of $H_N$. In other words, the maximum value of the bars under consideration is decisive to reject $F=G$. The weight makes, under $F=G$, the asymptotic variance of the bars independent of the positions in the plot. For more comments on weighted statistics in a two-sample setting, see Ledwina $\&$ Wy{\l}upek (2012a).

To apply ${\sf Max_{D(N)}}$ and to draw the related B-plot one has to decide about the form of $D(N)$. Through, we apply $D(N)$ to be the largest dimension not exceeding $N$. The choice is supported by existing evidence in the case of testing for stochastic dominance and the following result.\\

\noindent
{\bf Proposition 1.} {\it Suppose that $0 < \lambda_* \leq m/N\leq1-\lambda_* <1$, for some $\lambda_* \leq 1/2$. If $D(N)=o(N)$ and $D(N) \to \infty$ as $N \to \infty$ then ${\sf Max_{D(N)}}$ is consistent under any alternative to $F=G$.} \\

\noindent
{\sf 3.4. Real data examples}\\

We shall apply our procedures to data from the area of neuroscience, coming from the study by Talebi $\&$ Baker (2016). These data were already considered, in some parts, in Rousselet et al. (2017) and Wilcox $\&$ Rousselet (2023). Data were made available by the first mentioned researches and published online along with Rousselet et al. (2017).  The set of data has resulted from the recording and quantifying of the neurones inherent to the visual cortex of cats. In our description and discussion, we follow the notation and terminology introduced in Wilcox $\&$ Rousselet (2023). We consider three functionally distinct categories of simple cells: {\it nonoriented} (nonOri; 101 cases), {\it expansive oriented} (expOri; 48 cases), and {\it compressive oriented} (compOri; 63 cases), identified in the study by Talebi $\&$ Baker (2016). Four selected pairs of these outcomes were compared in Wilcox $\&$ Rousselet (2023) with respect to duration latencies and response latencies. 

We shall do an analogous exploration to this in Wilcox $\&$ Rousselet (2023), applying our approach to all six possible pairs of results. In Figure 1 we visualise these six pairs of samples via related B-plots. Moreover, along with each B-plot, local acceptance regions on level 0.95 are displayed, according to the description in Section 3.2. In addition, the related $p$-values of our global ${\sf Max_{D(N)}}$ test are given in each case. As said in Section 3.3, in each case we apply $D(N)$ to be the largest dimension not exceeding $N$. Consequently, $D(N)$ equals 127 in the first four cases and $D(N)=63$ in the last two situations depicted in the figure. The notation nonOri/expOri, etc., in the captions of the panels in Figure 1, means that nonOri corresponds to the first sample and therefore the population cdf $F$, while the results labelled expOri play the role of the second sample and obey the population cdf $G$, using our notation.

Wilcox and Rousselet (2023) begin visualisations in their Figure 7 by presenting, relevant to $F$ and $G$, density estimates in each of the categories and for the two latencies. 
This gives a preliminary insight into the differences between the compared distributions.  We present, in captions of panels in our Figure 1, $p$-values of our global ${\sf Max_{D(N)}}$ test, which is explicitly related to the B-plots presented there. This test reveals that, among response latencies, only differences between empirical distributions for nonOri and expOri, shown on pertinent B-plot, are not significant on the level 0.05, while, among duration latencies, differences between expOri and compOri are highly nonsignificant. The other four comparisons in our Figure 1 yield $p$-values being 0, in principle.

In terms of the shapes of the B-plots, at first sight,  Figure 1 suggests that in each of the six cases, we have a different pattern of mass allocation for the two samples under consideration. Moreover, the evident feature of all six B-plots is the dominating positivity of the bars. The interpretation of positivity is easy. In comparison $F/G$, given $p_0 \in (0,1),$ the bar value in $p_0$ is defined as $\eta_N {\widehat {\text {CCC}}}(p_0)$, i.e. equal to $w(p_0) \eta_N \bigl[F_m(H_N^{-1}(p_0))-G_n(H_N^{-1}(p_0))\bigr].$ The difference is positive if, below the $p_0$-quantile of the pooled sample, the observed frequency of observations from the first sample $(F)$ is higher than the related frequency in the second sample (G). In other words, in comparison to the first sample, in the second sample more observations exceed the $p_0$-quantile.  The value $w(p_0)$ rescales the difference in frequencies to provide a meaningful interpretation of the statement that the observed discrepancy is significantly large in the case of validity of ${\mathbb H}$. On an abstract level, the positivity of ${\text{CCC}}(p)$ for all $p \in (0,1)$ is equivalent to $F(x) \geq G(x),\; x \in {\mathbb R}$. This, in turn, means stochastic dominance of $G$ over $F$. 

Now, let us take a closer look at some of the panels and compare our conclusions with the related statements in Wilcox $\&$ Rousselet (2023). 

We start with a latencies comparison in the left upper panel. Local acceptance regions show that there is significant allocation of mass in the two first decile regions and in the last decile region. This indicates some statistically significant differences in the tails of $F$ and $G$. That is, the left tail of $G$ is thinner than that of $F$, while for the right tail the situation is reversed and the right tail of $G$ is fatter than the corresponding tail of $F$.  However, the observed differences in the tails (compared to the situation $F=G$) are not very big and they do not impact significantly into rejection of the global hypothesis $F=G$ at the standard level of 0.05. Regarding the findings of Wilcox $\&$ Rousselet (2023) in this case, the differences in the tails, which we have exhibited, are not seen on the plots of the estimated densities. Next, the graph of the Wilcox shift function in the second row of their Figure 7 suggests that the two groups differ a little for short latencies, and the difference increases as one moves from lower to upper deciles. However, no statistically significant differences between nine deciles in two groups have been found, on which they focus, using their approach. The 0.95 simultaneous confidence region, shown in Figure 7,  is rather wide, as typical for quantile-based solutions, and the uncertainty is large. 

The situation exemplified in the first panel, just discussed, seems to correspond to an abstract one in which ${\text{CCC}}(p) > 0$ on $(0,p_0)$, ${\text{CCC}}(p)=0$ on $[p_0,p_1]$, and ${\text{CCC}}(p) > 0$  on $(p_1,1)$. How to interpret such type shape of the CCC plot? Consider first the point $p_0$, which is the first $p,\;p >0,$ satisfying ${\text{CCC}}(p)=0$. Note also that always $F(H^{-1}(0))-G(H^{-1}(0))=0$. Hence, on the interval $[H^{-1}(0),H^{-1}(p_0)]$, both the distributions $F$ and $G$ locate the same amount of probability mass. Then, if for $p \in (0,p_0)$ we have ${\text {CCC}}(p) >0$, this means that, conditionally on the event: observations lay in $[H^{-1}(0),H^{-1}(p_0)]$, $F$ is stochastically smaller than $G$. The reverse inequality ${\text {CCC}}(p) < 0$ on $(0,p_0)$ would be equivalent to reversed (conditional) stochastic domination. If there are a few points $p$ such that ${\text {CCC}}(p)=0$, the interpretation extends to the related subintervals defined by the points that solve $F(H^{-1}(p))-G(H^{-1}(p))=0$. The case ${\text {CCC}}(p)=0$ for $p \in [p_0, p_1)$ is obvious. 

In the next case which we shall discuss and which is presented in the middle left panel in our Figure 1, we observe much better pronounced differences between the two samples. Now, significant local deviations are noticed in all decile regions, but the first one. So, no significant discrepancies are observed only in the left tail, which is related to quantiles such that $p \leq 0.1$. In turn, the most substantial allocation of mass is observed for the central quantiles region, that is, $p \in [0.4,0.6]$. In this case, our conclusions are to a greater extent consistent with those of Wilcox $\&$ Rousselet (2023). They have noticed that the deciles differ significantly except for the 0.1 quantile, while the magnitude of the differences between the deciles increases as we move from the lower to the upper deciles. 

We conclude our comments by considering the cell response durations and nonOri/expOri comparison. Our global ${\sf Max_{D(N)}}$ test evidently rejects $F=G$ and this is in agreement with the results of the investigations presented in Wilcox $\&$ Rousselet (2023). In addition, their finding on the increase in the differences in deciles beyond the median is consistent with the shape of our B-plot presented in the upper right panel in our Figure 1. However, it seems that this panel provides more detailed information on how the two samples differ.

To close the discussion, note that Brown and Zhang (2024) have also published recently a solution aimed at understanding the distributional differences between two samples. We have discussed extensively the binary expansion testing approach, which they exploit, in the context of independence testing in \'Cmiel and Ledwina (2024) and we skip related discussion and comparisons here.

\vspace{-0.5cm}
\begin{figure}[ht!]
\hspace{-2cm}
\includegraphics[trim = 10mm 50mm 10mm 20mm, clip, scale=1]{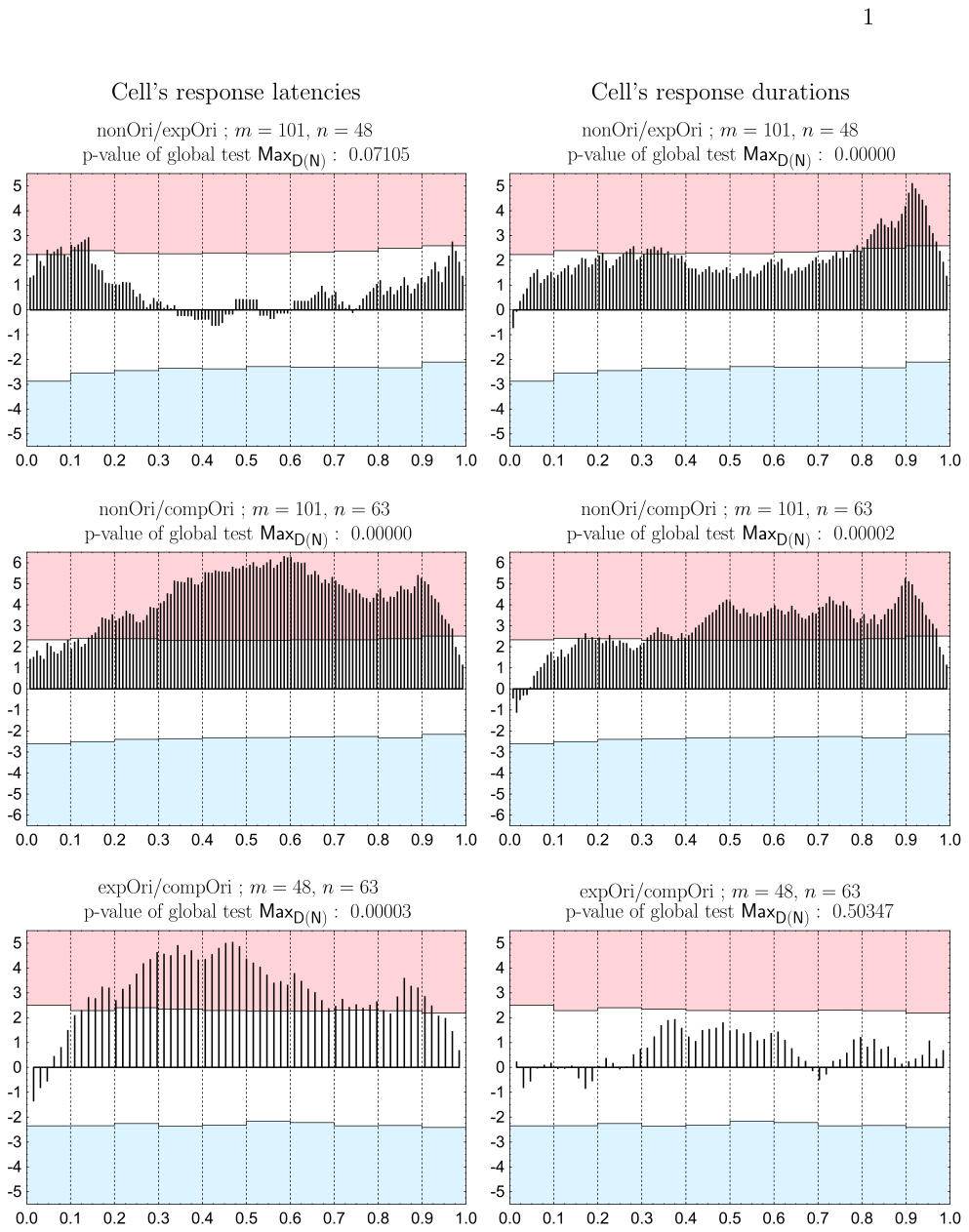}
\end{figure}

\noindent
{\bf Figure 1.} {\it Talebi-Baker data. Global and local comparisons of cells' response for the three categories: nonOri, expOri, and compOri. Left column: response latencies in ms; right column: response duration in ms.} \\
\newpage

\noindent
{\bf 4. Simulated powers of ${\sf Max_{D(N)}}$ and some of its competitors}\\

\noindent
{\sf 4.1. Compared tests}\\

Although many tests for $F=G$ against $F \neq G$ have been constructed, some new procedures still appear. We shall mention some of them, which are related either to the probability integral transformation or some relative distributions, and pertinent statistical functionals. As to the first group, it exploits the observation that if $F=G$ then $F(Y)$ should be uniformly distributed, where $Y \sim G$ and $X \sim F$. For some applications of this idea, see Nakas et al. (2003), Zhou et al. (2017),  Alonso et al. (2020) and Song $\&$ Xai (2022). Some procedures based on an empirical ROC curve were proposed by Pardo $\&$ Franco-Pereira (2017), Franco-Pereira et al. (2020) and Tang et al. (2021). Our first warning concerning this stream of works is that, in the case of unrestricted alternatives, the results of  such ``non-symmetric'' procedures may depend on numbering of the two samples. The second warning is considerable variability of the empirical ROC process; cf. our remarks in Appendix A.

Another stream of papers, related in turn to the process ${\widehat P}_N(p)$, includes Neuhaus (1987), Janic-Wr\'oblewska $\&$ Ledwina (2000), and Wy{\l}upek (2010), among others.

We shall compare the test ${\sf {Max_{D(N)}}}$ to convenient variant of integral Anderson-Darling statistics and to the solution $T_{new}$ of Tang et al. (2021). The last mentioned paper matches some transformation of data with parametric modeling. This is a complex construction. We refer to Tang et al. (2021) for details. In our simulations, we have applied this test, including the modification introduced in their codes in R. We shall abbreviate the notation $T_{new}$ to ${\sf T_N}$. As to the integral Anderson-Darling test in the two-sample setting, the most popular variant is Pettitt's (1976) construction; see also Scholz $\&$ Stephens (1987) for further development. It has the form
$$
\eta_N \left\{\int\limits_{\{z: H_N(z)<1\}} \left[\frac{F_m(z)-G_n(z)}{\sqrt{H_N(z)(1-H_N(z))}}\right]^2 dH_N(z)\right\}^{1/2}.
\eqno(7)
$$
In our simulations we shall relay on (2) and (4) and we shall use the following, slightly simpler, variant of this solution
$$
{\sf AD_N}=\eta_N \left\{\int\limits_0^1 \left[\frac{F_m(H_N^{-1}(p))-G_n(H_N^{-1}(p))}{\sqrt{p(1-p)}}\right]^2 dp\right\}^{1/2}=
\eta_N \left\{\int\limits_0^1 \left [ {\widehat {\text {CCC}}}(p) \right]^2dp\right \}^{1/2}=
$$
$$
\left\{\int\limits_0^1 \left [ {\sf P}_N(p) \right]^2dp\right \}^{1/2}=
\eta_N \left\{\sum\limits_{k=1}^{N-1} \left[\frac{S_k}{m}-\frac{T_k}{n}\right]^2 \log\left(\frac{(k+1)(N-k+1)}{k(N-k)}\right) \right\}^{1/2},
$$
where $S_k$ and $T_k$ are the relative ranks. More precisely, $S_k=mF_m(H_N^{-1}(\frac{k}{N}))$ and $T_k=nG_n(H_N^{-1}(\frac{k}{N}))$, $k=1,...,N$.
As $\;0 \leq H_N(H_N^{-1}(p)) - p \leq 1/N$,   the variant ${\sf AD_N}$ is very close to (7), but explicitly related to ${\widehat {\text {CCC}}}(p)$. We find it to be useful for interpreting the empirical results in terms of the allocation of mass exhibited by the CCC. In our study, simulated powers of both variants have differed no more than by 0.01. Therefore, we present the results for ${\sf AD_N}$, only.\\

\noindent
{\sf 4.2. Alternative models and empirical powers}\\

We have considered several alternative models pertinent to standard cdf's $F$ and $G$. In an extensive simulation study of Tang et al. (2021) they have restricted attention to three location-scale models and have enquired on the influence of shift, scale, and both parameters on empirical powers. In such a framework, they have got very encouraging results about ${\sf T_N}$. We have included such location-scale models to our study as well, but we have also tried some others already considered in the literature. Our list of models is given in Table 1 and we display related CCC's in Figure 2. The CCC curves were obtained by averaging 1000 empirical ${\text{CCC}}(\cdot)$ calculated for $m=n=5 000$ observations from $F$ and $G$, respectively. When calculating empirical powers, we have considered $\alpha=0.05,\;m=n=100$, $D(N)=127$, and we have done 10 000 MC repetitions in each case. The critical values of ${\sf {Max_{D(N)}}}$ and ${\sf {AD_N}}$ were obtained using 100 000 MC runs and are equal to: 2.9933 and 1.5687, respectively.  The solution ${\sf T_N}$ requires an application of bootstrap. We have written C++ codes for all computations in our paper.

Closely related, a one-sided variant $M_{D(N)}$ of ${\sf {Max_{D(N)}}}$, has been extensively studied in Ledwina $\&$ Wy{\l}upek (2012a, 2013) and in Inglot et al. (2019), and appears to be a stable and powerful solution, detecting a wide spectrum of one-sided alternatives with high frequency. In view of this evidence, we have treated the present statistic ${\sf {Max_{D(N)}}}$  as a benchmark, and we have selected the parameters for the alternatives in such a way that the empirical powers of ${\sf {Max_{D(N)}}}$  were in $[0.75,0.80]$.  Next, we have sorted the alternative models under consideration according to the decreasing empirical powers of ${\sf {AD_N}}$, which resulted in the list ${\mathbb A}_1 - {\mathbb A}_{18}$, appearing in Table 1 and Figure 2. Since, up to $\eta_N$, the statistic ${\sf {AD_N}}$ estimates the $L_2$ norm of CCC, looking at Figure 2, the result of the sorting can be guessed quite accurately, without looking at the pertinent empirical powers. In addition, the empirical powers of ${\sf {AD_N}}$ are consistent with the existing evidence on this solution. In particular, ${\sf {AD_N}}$ is highly powerful in detecting considerable allocation of probability mass to central range of quantiles, as well as towards moderate range of quantiles. When most of the significant changes occur near 0 or 1 ${\sf {AD_N}}$ loses part of its high power.

Our simulations have shown that ${\sf T_N}$ is not very stable if one considers more complex alternatives than those investigated by Tang et al. (2021). Although in many cases ${\sf T_N}$ characterises high power, in some others its powers are not impressive. Understanding the weaknesses of ${\sf T_N}$ would require additional consideration.

${\sf {Max_{D(N)}}}$ proves to be a simple, intuitive, and powerful solution that provides stable and high power over a variety of alternatives.\\
\newpage
\noindent
{\bf Table 1.} {\it Empirical powers of $\;{\sf Max_{D(N)}}$,  ${\sf AD_N}$, and  ${\sf T_N}$ under alternatives} ${\mathbb A}_1 - {\mathbb A}_{18}$; $\alpha=0.05$, $m=n=100$, $D(N)=127$.\\

\begin{table}[ht!]
\begin{tabular}{c c c c c c}
 Notation         & Description & ${\sf Max_{D(N)}}$ & ${\sf AD_N}$ & ${\sf T_N}$ \\
\\
\hline
\\
$\mathbb{A}_1$    &$N(0,1)/N(0.45,1)$ &	  76      &   86     &   79        \\ 
$\mathbb{A}_2$    &$\mathrm{Pareto}(1)/\mathrm{Pareto}(1.6)$ &	  77      &   84     &   74        \\ 
$\mathbb{A}_3$    &$\mathrm{Laplace}(0,1)/\mathrm{Laplace}(0.4,1.6)$ &	  75       &   79     &   92        \\ 
$\mathbb{A}_4$    &$U[0;1]/[(0.58)U[0,1]+(0.42)\mathrm{Beta}(50,50)]$ &	  78       &   79     &   77        \\ 
$\mathbb{A}_5$    &$LN(0.92,0.5)/LN(1.08,0.4)$ &	  77        &   77     &   69        \\ 
$\mathbb{A}_6$    &$N(0,1)/[(0.6)N(-0.9,0.37)+(0.4)N(1,0.7)]$  &	  76      &   73     &    31       \\
$\mathbb{A}_7$    &$N(0,1)/N(0,1.55)$ &	  75      &   69     &   96        \\ 
$\mathbb{A}_8$    &$\mathrm{Fan}(0.66)/ \mathrm{Uniform}(-1,1)$ &	  75       &   63     &   25        \\ 
$\mathbb{A}_9$    &$N(0,1)/\mathrm{Anderson}(1.5)$ &	  77       &   61     &   13        \\ 
$\mathbb{A}_{10}$ &$[(0.52)N(0.4,1)+(0.48)\chi^2_1]/N(0.4,1)$ &	  79        &   59     &   53        \\
$\mathbb{A}_{11}$ &$N(0,1)/[(0.8)N(0,1)+(0.2)\mathrm{Lehmann}(0.16)]$   &	  75      &   57     &    73       \\ 
$\mathbb{A}_{12}$ &$LN(0,1)/LNC(1,1.8)$ &	  77       &   52     &   80        \\ 
$\mathbb{A}_{13}$ &$\mathrm{Lehmann}(1.2)/\mathrm{Subbotin}(8)$  &	  75     &   41     &    26       \\ 
$\mathbb{A}_{14}$ &$N(0,1)/[(0.35)N(0,1)+(0.65)\mathrm{Cauchy}(0,1)]$ &	  75        &   38     &   79        \\ 
$\mathbb{A}_{15}$ &$\mathrm{Exp}(1)/[\mathrm{Exp}(1)+0.11]$ &	  75        &   38     &   24        \\ 
$\mathbb{A}_{16}$ &$N(1.7,1.7)/\mathrm{Gamma}(1.7,1)$  &	  75     &   38     &    11       \\ 
$\mathbb{A}_{17}$ &$N(0,1)/\mathrm{Cauchy}(0,0.7)$ &	  79     &   26     &   54        \\ 
$\mathbb{A}_{18}$ &$[U(0,1)]/\mathrm{Mason-Schuenemeyer}(20,0.1)$ &	  75       &   11     &   9         \\
\\
\hline
Mean              & &   76  &  57        &   54

\end{tabular}
\end{table}

In Table 1, we have described each of the alternatives by the symbol $F/G$, where $F$ and $G$ are some known continuous cdf's. The abbreviation $LN(\mu,\sigma)$ denotes the log-normal distribution with parameters $\mu$ and $\sigma$. Fan($\theta)$ stands for the local departure model proposed in Fan (1996). Anderson$(\theta)$ denotes the kurtotic distribution generated as $X|X|^{\theta},\;X \sim N(0,1),\;\theta \geq 0$.
By Lehmann$(\theta)$ we mean cdf $[\Phi(x)]^{\theta},\;\theta \geq 0,$ where $\Phi(x)$ is the cdf of $N(0,1)$. $LNC(\sigma_1,\sigma_2)$ is a symbol of two-piece log-normal distribution introduced in Ledwina $\&$ Wy{\l}upek (2012a). Finally, Mason-Schuenemeyer$(\beta,\theta)$ stands for the tail distribution family introduced in Mason $\&$ Schuenemeyer (1983). Other notations are standard and are therefore introduced in a not very formal way.

\newpage

\begin{figure}[ht!]
\hspace{0.5cm}
\includegraphics[trim = 25mm 10mm 20mm 3mm, clip, scale=0.90]{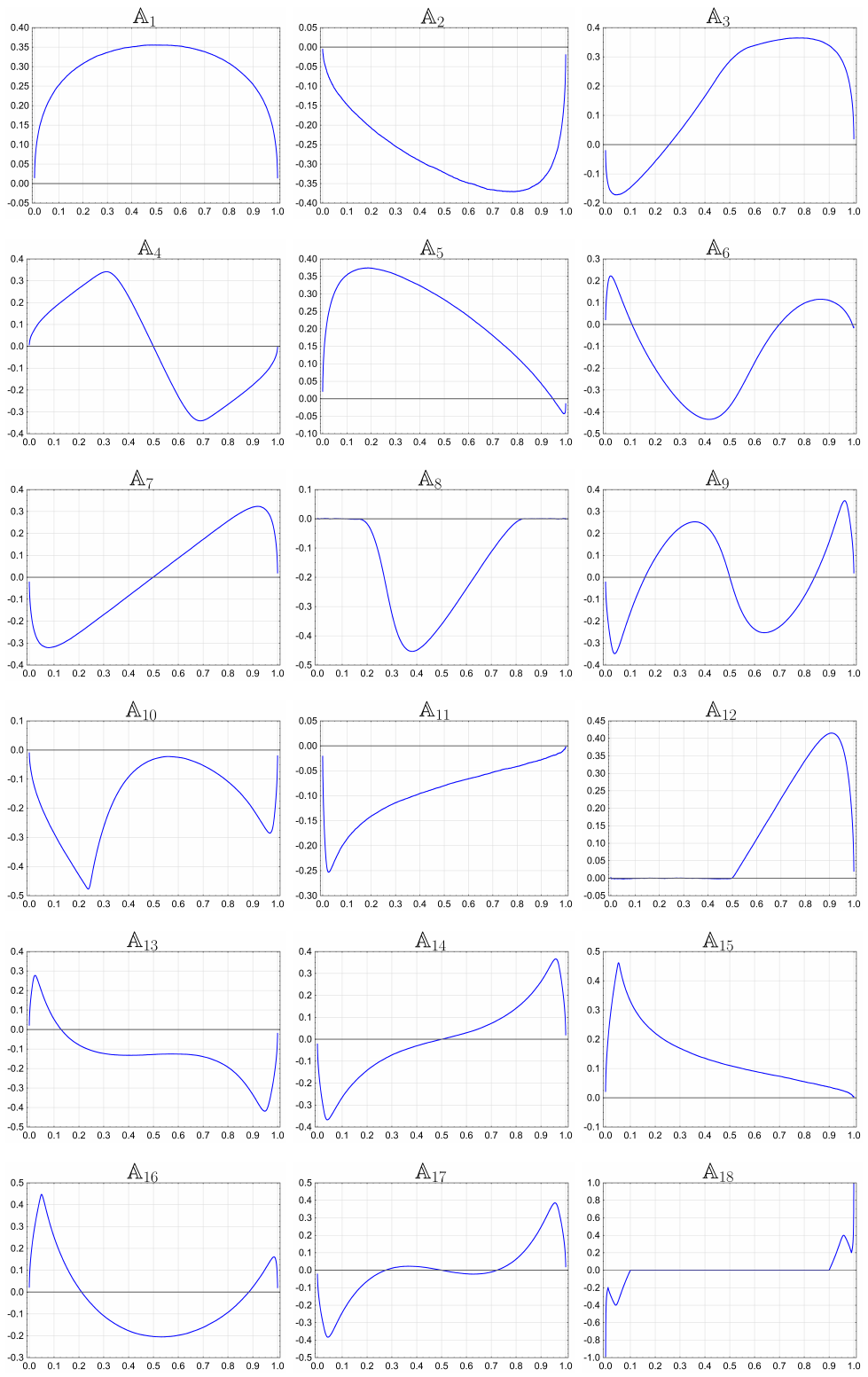}
\end{figure}
\vspace{-0.8cm}
{\bf Figure 2.} {\it Graphs of} CCC {\it curves for alternatives} ${\mathbb A}_1 - {\mathbb A}_{18}$; $\lambda_N=0.5$. {\it The curves are shown in the interval} [0.0001,0.9999] {\it in each case.}\\

\noindent
{\bf 5. Discussion}\\

This paper introduces a new methodology for the comparison of two samples and demonstrates its practical benefits. Our approach offers valuable new tools for exploring and understanding distributional differences between two groups of data. B-plots along with local acceptance regions allow for identifying specific quantile ranges where the differences are statistically significant. Both simulation results and real-world data examples show the ability of the approach to reveal a wide range of discrepancies between distributions. The proposed tools are computationally efficient and highly useful in visual interpreting and analysing differences between samples.

Data driven Neyman's tests by Janic-Wr\'oblewska $\&$ Ledwina (2000) and Wy{\l}upek (2010) were presumably first two-sample tests associated with some discussion on structure of underlying models when testing $F=G$, against unrestricted alternative $F \neq G$. An attempt was realised by calculating, averaging, and displying smooth components, or equivalently, adequate Fourier coefficients pertinent to the current alternative. However, the components in these tests were based on classical Legendre polynomials and therefore an intuitive interpretation of departures was restricted to two or three first components. Ledwina $\&$ Wy{\l}upek (2012a) have constructed, again in a two-sample setting, a data-driven test with nonstandard components, allowing for much more insightful presentation and interpretation of data at hand. These components appeared to be handy to understand, in case of rejection of $F=G$, 
where and how two samples differ.  These nonstandard components coincide, up to continuity correction and completely different notation, with presently introduced bars. The construction of this new data-driven test was rather complex. The resulting empirical powers were very satisfactory.
In the same paper by Ledwina $\&$ Wy{\l}upek (2012a), minimum of the components was also tried, as a complementary way of combining them. The resulting test for $F=G$ against $F \not\ge G$, in extensive simulations, appears to be as good as the much more complex data-driven test. 

In view of the above, in the present paper we have restricted our attention to ${\sf {Max_{D(N)}}}$, only. On the other hand, when testing Gaussianity, a counterpart of ${\sf {Max_{D(N)}}}$ has some weaknesses, and Ducharme and Ledwina (2024) have proposed a much better data-driven test based on bars pertinent to this problem. In turn, an analogue of ${\sf {Max_{D(N)}}}$ for testing uniformity works very well, while similar construction in the independence testing problem requires some smoothing; cf. \'Cmiel et al. (2020) and \'Cmiel $\&$ Ledwina (2024). This evidence shows that, in general, it is difficult to make a general recommendation about which of these two methods is the best. Our conclusion is that the two-sample process ${\sf P}_N(p)$ is specific and allows for a simple and powerful construction. 

The last group of our comments is on test comparison. In our simulation study, we have considered a longer list of models than finally presented in Table 1 and Figure 2. It is not surprising that analytically different pairs $F/G$ can yield a very similar mass allocation expressed by related CCC curves. Therefore, using CCC plots, we eliminated several cases and selected those that were clearly different. Note also that CCC$(p)$ represent aggregated Fourier coefficients of an appropriate comparison density in the system of projected Haar functions; cf. Ledwina $\&$ Zagda\'nski (2024) for details. 

A recent paper by Kodalci $\&$ Thas (2024) nicely describes the evident chaos noticed in many simulation studies on the evaluation of two-sample tests. These authors also propose a new open science initiative for neutral comparison of tests to verify $F=G$. The number of numerical comparisons they offer seems to be almost infinite. In this context, we would like to propose looking at the issue from a different perspective and returning to classical methods of test evaluation by comparing efficiency. Inglot et al. (2019) have elaborated a relatively simple approach, the so-called pathwise intermediate efficiency, which is applicable in standard nonparametric testing problems. It gives useful insight into structure and abilities of statistical functionals, being used to define tests statistics, and allows to understand related expected finite sample tests behaviour for a reasonable range of relatively small significance levels and wide range of sample sizes.  The illustrative comparisons done in Inglot et al. (2019) for some two-sample tests confirm this statement. See also \'Cmiel et al. (2020) as an example in the case of testing the simple null goodness-of-fit hypothesis. \\

\noindent
{\bf Appendix A: Contrasting two-rank processes under F=G}\\

In this section, we shall compare finite sample behavior of two processes 
$$
{\widehat P}_N(p)=\eta_N [F_m(H_N^{-1}(p))-G_n(H_N^{-1}(p))]\;\;\mbox{and}\;\;{\widehat U}_N(p)=\eta_N [p-G_n(F_m^{-1}(p))],\;\;p \in [0,1],
$$
under $F=G$. From existing results, collected in Appendix A in Ledwina $\&$ Zagda\'nski (2024), it follows that, under $F=G$, the asymptotic behaviour of both processes is the same. However, in extensive simulations presented in this paper,  procedures based on ${\widehat U}_N(p)$ have exhibited greater variability, in particular under $F=G$, than analogous objects based on ${\widehat P}_N(p)$. The goal of this section is to attempt to understand a source of problems of this type. In view of the great popularity of the empirical ROC process, clearly related to ${\widehat U}_N$, the question seems to be of independent interest. 

We start with auxiliary notation and alternative formulas for the two processes. Let $X_{1:m},...,X_{m:m}$ and $Y_{1:n},...,Y_{n:n}$ denote ordered values in the first and second samples, respectively. Analogously, $Z_{1:N},...,Z_{N:N}$ stands for ordered values in the pooled sample $(X_1,...,X_m,Y_1,...,Y_n)$. 

Now, let $R_k,\; k=1,...,m,$ denote the rank of $X_{k:m}$ in the pooled sample. Then $nG_n(X_{k:m})=R_k -k,\;k=1,...,m$. Hence, any realization of the process ${\widehat U}_N$ can be expressed as 
$$
{\widehat U}_N(p)=\eta_N \sum_{k=1}^m\bigl[p-G_n(X_{k:m})\bigr]\mathbbm{1}\bigl(\frac{k-1}{m} < p \leq \frac{k}{m} \bigr)=
$$
$$
\eta_N \sum_{k=1}^m\bigl[p-\frac{R_k -k}{m}\bigr]\mathbbm{1}\bigl(\frac{k-1}{m} < p \leq \frac{k}{m} \bigr),
\eqno({\text{A1}})
$$
where $\mathbbm{1}(E)$ denotes the indicator of the event $E$. Note also that $F_m^{-1}(0)$ is set to $X_{1:m}$, as commonly set, which completes this formula for the case $p=0$.

Next, let $S_i,\;i=1,...,N,$ be the number of observations among $X_1,...,X_m$ that do not exceed $Z_{i:N}$. Then, $S_i =mF_m(H_N^{-1}(\frac{i}{N})),\;i=1,...,N.$ We also introduce
$T_i =nG_n(H_N^{-1}(\frac{i}{N})),\;i=1,...,N,$ to denote analogously defined ranks of $Y_1,...,Y_n$. With these notation, 
$$
{\widehat P}_N(p)=\eta_N \sum_{i=1}^N\bigl[F_m((H_N^{-1}(p))-G_n(H_N^{-1}(p))\bigr]\mathbbm{1}\bigl(\frac{i-1}{N} < p \leq \frac{i}{N} \bigr)=
$$
$$
\eta_N \sum_{i=1}^N\bigl[\frac{S_i}{m}-\frac{T_i}{n}\bigr]\mathbbm{1}\bigl(\frac{i-1}{N} < p \leq \frac{i}{N} \bigr).
\eqno({\text{A2}})
$$
Again, we additionally set $H_N^{-1}(0)=Z_{1:N}$. The above introduced ranks $R_k$'s, $S_i$'s and $T_i$'s are termed in the pertinent literature as relative ranks.

\begin{figure}[ht!]
\hspace{-1.4cm}
\includegraphics[trim = 10mm 120mm 10mm 25mm, clip, scale=1]{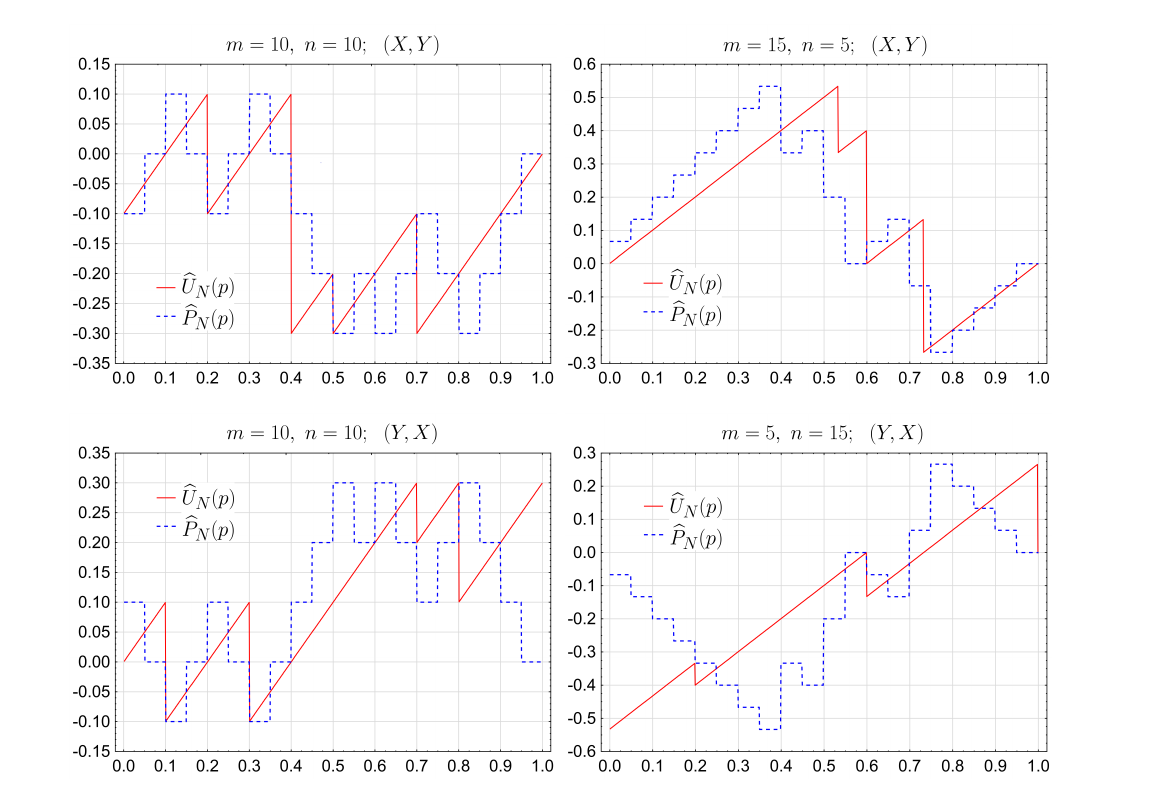}
\end{figure}
\noindent
{\bf Figure A1. } {\it Realisations of processes ${\widehat U}_N(p)$ and ${\widehat P}_N(p)$ with equal and unequal sample sizes; first row. Realisations of ${\widehat U}_N(p)$ and ${\widehat P}_N(p)$ for renumbered samples; second row.}\\

The first evident observation from the formulas $({\text{A1}})$ and $({\text{A2}})$ is that the number of values of the rank process ${\widehat P}_N(p)$ is $m+n$, while the number of values of the rank process ${\widehat U}_N(p)$ is $m$. So, ${\widehat P}_N(p)$ provides a much more careful inspection over time $p$. The next simple but useful observation is that the renumbering of two samples transforms ${\widehat P}_N(p)$ to $-{\widehat P}_N(p)$. However, in the case of the process ${\widehat U}_N(p)$ such an operation yields very different, in structure, realisations, as a rule. For an illustration, in Fig. A1 we present realisations of the two processes under two choices of small sample sizes and two configurations of samples : (X,Y) and (Y,X). Both samples $X_1,...,X_m$ and $Y_1,...,Y_n$ were generated from the same continuous cdf.

To study the variability of both processes through a formal argument, below we give the expected values and variances of ${\widehat U}_N(p)$ and ${\widehat P}_N(p)$,  under $F=G$ and fixed $m$ and $n$. Recall that asymptotically, under $F=G$, these functions are identical and equal to 0 and $p(1-p)$, respectively. \\

\noindent
{\bf Proposition 2.} {\it If $F$ is a continuous cdf and $G=F$ then}
$$
\text{E}({\widehat U}_N(p))=\eta_N \Bigl(p-\frac{\lceil pm \rceil}{m+1}\Bigr)\;\;\;{and}\;\;\; \text{E}({\widehat P}_N(p))=0,
$$
{\it while}
$$
\text{Var} ({\widehat U}_N(p))=\eta_N^{2}\frac{\lceil pm \rceil(m-\lceil pm \rceil+1)(N+1)}{(m+1)^2(m+2)n} \;\;\;{and}\;\;\; \text{Var} ({\widehat P}_N(p))=\eta_N^{2}\frac{\lceil pN \rceil(N-\lceil pN \rceil)}{mn(N-1)}.
$$
\\
A verification of these formulas is postponed to Appendix B.  \\

In Figure A2, we give a numerical illustration of the shapes of $\text{Var} ({\widehat U}_N(p))$ and $\text{Var} ({\widehat P}_N(p))$ under two selections of small sample sizes $m$ and $n$. The evidence is supplemented by graphs of the function 
$$
\Delta_N(p)=\{\text{Var}({\widehat U}_N(p)) - \text{Var}({\widehat P}_N(p))\}/\{p(1-p)\}, 
$$ 
which shows  how variance functions of the two processes differ relative to the asymptotic variance $p(1-p)$. This illustration shows that ${\widehat U}_N(p)$ has a smaller exact variance function than ${\widehat P}_N(p)$ for centrally located $p$'s. On the other hand, for $p$'s close to 0 and 1 the exact variance function of ${\widehat U}_N(p)$ is considerably greater than that of ${\widehat P}_N(p)$. Consequently, we conclude that the exact variance function of ${\widehat P}_N(p)$ is much closer to the asymptotic one than the exact variance function of ${\widehat U}_N(p)$. The above reveals greater stability, under finite sample sizes and $F=G$,  of the process ${\widehat P}_N(p)$ compared to ${\widehat U}_N(p)$.
\newpage
\begin{figure}[ht!]
\hspace{-1.25cm}
\includegraphics[trim = 10mm 40mm 25mm 30mm, clip, scale=1]{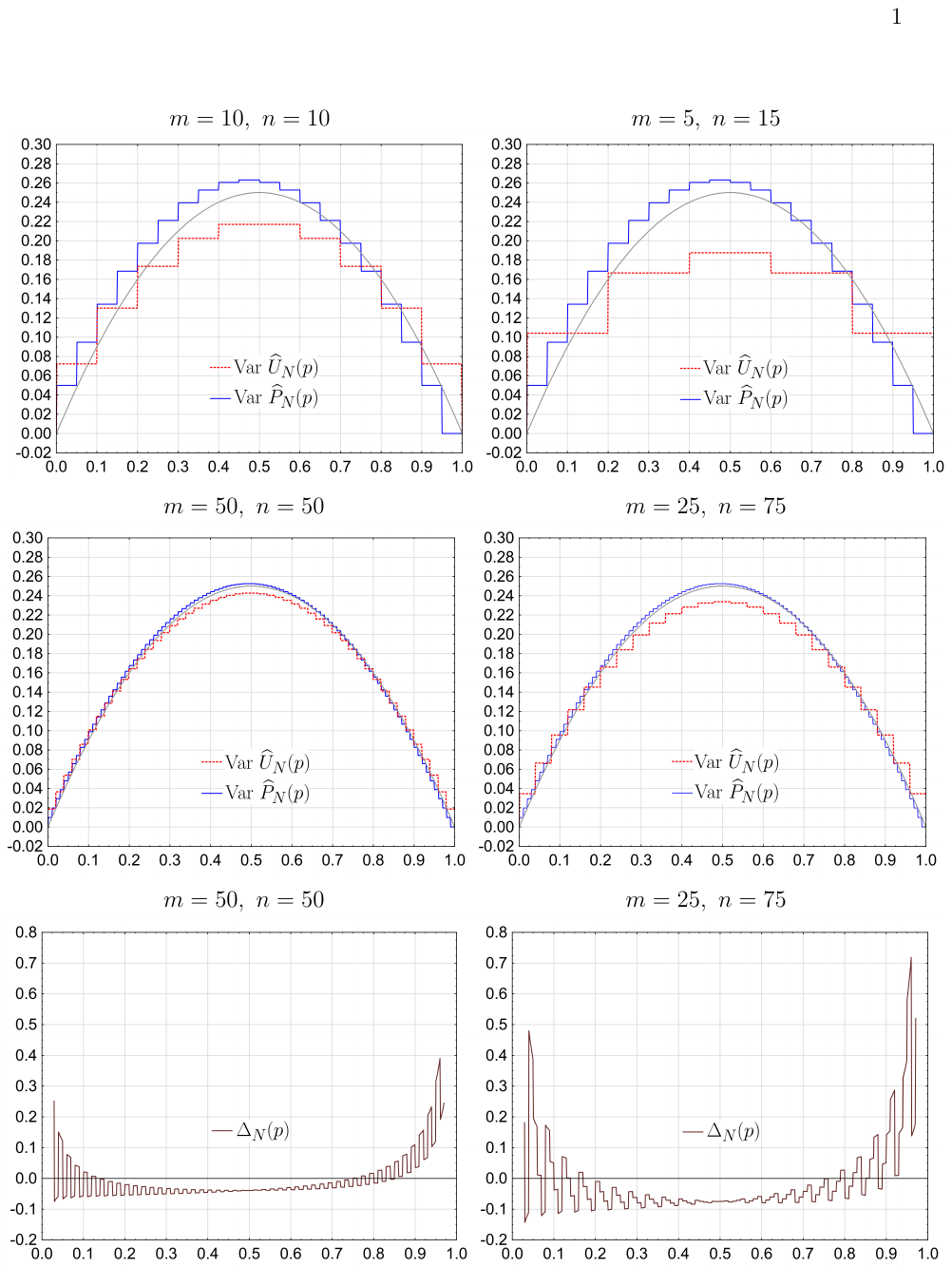}
\end{figure}
\noindent
{\bf Figure A2.} {\it Comparison of variances under $F=G\;$: ${\mathrm Var} ({\widehat U}_N(p))$ to  ${\mathrm Var} ({\widehat P}_N(p))$,  against selected sample sizes; first and second row. The light grey parabola represents the joint asymptotic variance of both processes. The third row shows $\Delta_N(p)$ against selected sample sizes. The graph of $\Delta_N(p)$ is shown in the interval} [0.03,0.97].
\\
\newpage
In the context of the above results for the case $F=G$, recall again  that under $F \neq G$  greater variability of realizations of ${\widehat U}_N(p)$ is expected, as a rule. For related results, see Appendix A in Ledwina $\&$ Zagda\'nski (2024).

In conclusion, the facts collected above indicate that it is much easier to construct some stable statistics based on ${\widehat P}_N(p)$ than their counterparts based on the process ${\widehat U}_N(p)$, relevant to the empirical ROC curve. However, the clear and easy interpretability of the results is retained in both approaches. This accounts for the advocacy to use the process ${\widehat P}_N(p)$ more widely than until now. \\

\noindent
{\bf Appendix B: Proofs}\\

\noindent
{\bf Proof of Proposition 1.} Our proof exploits Lemma 3 of Ledwina $\&$ Wy{\l}upek (2012b) and is to some extent similar to the proof of Lemma 2 in Ledwina $\&$ Wy{\l}upek (2012a). However, we avoid an application of Theorem 4.1 from Pyke and Shorack (1968), and, in this way, we skip the additional assumption appearing in Lemma 2 of Ledwina $\&$ Wy{\l}upek (2012a). Moreover, the statistic $M_d$, appearing in this Lemma 2, is based on some linear rank statistics with the correction for continuity inserted. This produces some negligible but noisy terms in comparison to the present setting. Finally, many of the notation of the present paper are different from those in Ledwina $\&$ Wy{\l}upek (2012a,b). This makes it difficult to follow these materials. Therefore, for the convenience of the reader, we provide some details.

To be specific, let ${\cal R}_i,\;i=1,...,m$ denote the rank of $X_i$ in the pooled sample $X_1,...,X_m,\\Y_1,...,Y_n$. Similarly, ${\cal R}_i,\;i=m+1,...,N$, stands for the rank $Y_i$ in the pooled sample.
Given $j=1,...,D(N)$, set 
$$
{\ell}_j(p) = {\ell}_{jN}(p)=- \sqrt{\frac{1-p_{S(N),j}}{p_{S(N),j}}}\, \mathbbm{1}(0 \leq p < p_{S(N),j}) +
\sqrt{\frac{p_{S(N),j}}{1- p_{S(N),j}}}\, \mathbbm{1}(p_{S(N),j} \leq p \leq 1).
$$

Consider the corresponding linear rank statistics given by 
$$
{\cal L}_j ={\cal L}_{jN}= \sum_{i=1}^{N} c_{Ni}\,{\ell}_j\Bigl(\frac{R_i-0.5}{N}\Bigr),\;\;\;j=1,...,D(N),
$$
where
$$
c_{Ni} = \sqrt{\frac{mn}{N}}
\left\{\begin{array}{lrl}
- m^{-1} & \mbox{if} &  1\leq i \leq m, \\
\quad n^{-1} & \mbox{if} &  m <   i \leq N.
\end{array} \right.
$$
Additionally, introduce
$$
{\cal T}_N = \max_{1 \leq j \leq D(N)} \bigl\{-{\cal L}_j\bigr\}.
$$
It holds that
$$
|{\cal T}_N - {\tt Max_{D(N)}}| \leq C_{\lambda_*} /\sqrt {N \min\{p_{S(N),j},1-p_{S(N),j}\}},
$$
where $C_{\lambda_*}$ is a positive number which depends only on $\lambda_*$. Hence, if $D(N)=o(N)$, then
$$
|{\cal T}_N - {\tt Max_{D(N)}}| =o(1).
$$
On the other hand, by Lemma 3 of Ledwina $\&$ Wy{\l}upek (2012b), under $F=G$, it holds that
$$
{\cal T}_N =O_P(\sqrt{\log \log D(N)}).
$$
By the above, the same relation holds for ${\sf Max_{D(N)}}$. Hence, the critical value of the test based on ${\sf Max_{D(N)}}$, say $c(\alpha,N)$, is of the order $O(\sqrt{\log \log D(N)})$.

In the case  $F \neq G$, consider first the process $\widehat {P}_N(p)$, related to the process ${\sf P}_N(p)$ via re-weighting, and introduce
$$
\nabla_N(p)=F(H_N^{-1}(p))-G(H_N^{-1}(p)):=(F-G) \circ H_N^{-1}(p)\;\;\;\mbox{and}\;\;\;\nabla(p)=(F-G) \circ H^{-1}(p). 
$$
With these notation
$$
\widehat {P}_N(p)=K_N(p)  + \eta_N \nabla_N(p),\;\;\;\mbox{where}\;\;\;K_N(p)=\eta_N\Bigl[(F_m-G_n) \circ H_N^{-1}(p) - \nabla_N(p)\Bigr].
$$
By our assumption, $\eta_N=O(\sqrt N)$. Hence, by weak convergence of two-sample Kolmogorov-Smirnov statistic, the  term $K_N(p)$ is $O_P(1)$.

If $F \neq G$, then there exists $j_0$, $j_0$ independent of $N$, such that $\nabla(p_{S(N),j_0}) \neq 0$, for all $N$ sufficiently large. For this $j_0$ write 
$$
P\Bigl({\tt Max_{D(N)}}\geq c(\alpha,N)\Bigr) \geq P\Bigl(w(p_{S(N),j_0})\bigl|{K_N}(p_{S(N),j_0}) +\eta_N \nabla_N(p_{S(N),j_0})\bigr| \geq c(\alpha,N)\Bigr).
$$
Since $\eta_N \nabla_N (p_{S(N),j_0})=O(\sqrt N)$ while $D(N)=o(N)$, therefore Proposition 1 is proved. \hfill $\Box$\\

\noindent
{\bf Proof of Proposition 2.} Under $F=G$ both processes are distribution-free. Therefore, without loss of generality, in this proof we assume that all observations $X_1,...,X_m$ and  $Y_1,...,Y_n$ obey $U(0,1)$ law. This, in particular,  implies that $X_{k:m} \sim \mathrm{Beta}(k,m-k+1)$. This fact is exploited in the case of the process $\hat {U}_N(p)$.

It holds that
$$
\eta_N^{-1}{\text E}({\widehat U}_N(p))=\sum_{k=1}^m  {\text E}\left[p-G_n(X_{k:m})\right]\mathbbm{1}\left(\frac{k-1}{m} < p \leq \frac{k}{m} \right).
$$
Moreover
$$ {\text E}\left[p-G_n(X_{k:m})\right]= {\text E}\left[{\text E}\left(p-G_n(X_{k:m})|X_{k:m}\right)\right]={\text E}\left[p-\frac{1}{n} n X_{k:m}\right]=p-\frac{k}{m+1} $$
and 
$${\text E}\left[p-G_n(X_{k:m})\right]^2={\text E}\left[{\text E}\{(p-G_n(X_{k:m})\}^2|X_{k:m}\right]$$
$$=p^2-2p\frac{k}{m+1}+\frac{1}{n^2}{\text E}\left[ n X_{k:m}(1-X_{k:m}) + n^2 (X_{k:m})^2\right]$$
Known formulas for the expectation and the variance of the $\mathrm{Beta}(k,m-k+1)$  distribution yield the result.\\

In the case of the process $\hat {P}_N(p)$ we shall in turn use the observation that, under $F=G$, the relative rank $S_k$ has the Hypergeomertic distribution with parameters $(N,k,m)$. Hence
$${\text {E}}(S_k)=\frac{mk}{N}, \ \ \ {\text{Var}}(S_k)={\text E} \left[S_k-\frac{mk}{N}\right]^2=\frac{mk(N-k)(N-m)}{N^2(N-1)}.$$
On the other hand, by (4), it holds that
$$
{\widehat P}_N(p)=\eta_N \sum_{k=1}^N  \left[\frac{S_k}{m}-\frac{k-S_k}{n}\right]\mathbbm{1}\left(\frac{k-1}{N} < p \leq \frac{k}{N} \right)$$
$$
=\eta_N \sum_{k=1}^N  \frac{N}{mn}\left[S_k-\frac{mk}{N}\right]\mathbbm{1}\left(\frac{k-1}{N} < p \leq \frac{k}{N} \right).
$$
By the above, the conclusions follow. \hfill $\Box$\\

\noindent
{\bf References}
\\
\setlist{nolistsep} 
\begin{spacing}{1}  
\begin{description}[topsep=0pt,itemsep=0pt,parsep=0pt,labelsep=0em]	
\item
Alonso, R., Nakas, C.T. $\&$ Carmen Pardo, M. (2020). A study of indices useful for the assessment of diagnostic markers in non-parametric ROC curve analysis. {\it Communications in Statistics-Simulation and Computation} {\bf 49}, 2102-2113.

\item
Anderson, G. (1996). Nonparametric tests of stochastic dominance in income distributions. {\it Econometrica} {\bf 64}, 1183-1193.

\item
Behnen, K. $\&$  Neuhaus, G. (1983). Galton's test as a linear rank test with estimated scores and its local asymptotic efficiency. {\it Annals of Statistics} {\bf 11}, 588-599.

\item
Borovkov, A.A., $\&$ Sycheva, N.M. (1968). On asymptotically optimal non-parametric criteria. {\it Theory of Probability and Its Applications} {\bf 13}, 359-393.

\item
Brown, B. $\&$ Zhang, K. (2024). AUGUST: An interpretable, resolution-based two-sample test. {\it New England Journal of Statistics in Data Science} {\bf 2}, 357-367.

\item
Campbell, G. (1994). Advances in statistical methodology for the evaluation of diagnostic and laboratory tests. {\it Statistics in Medicine} {\bf 13}, 499-508.

\item
\'Cmiel, B., Inglot, T. $\&$ Ledwina, T. (2020). Intermediate efficiency of some weighted goodness-of-fit statistics. {\it Journal of Nonparametric Statistics} {\bf 32}, 667-703.

\item
\'Cmiel, B. $\&$ Ledwina, T. (2024). Detecting dependence structure: visualization and inference. {\it arXiv} : 2410.05858v1 [stat.Me]

\item
De Jong, R., Liang, C.C. $\&$  Lauber, E. (1994). Conditional and unconditional automaticity: A dual-process model of effects of spatial stimulus-response concordance. {\it Journal of Experimental Psychology: Human Perception and Performance} {\bf 20}, 731-750.

\item
Doksum, K.A. $\&$  Sievers, G.L. (1976). Plotting with confidence: Graphical comparisons of two populations. {\it Biometrika} {\bf 63}, 421-434.

\item
Duong, T. (2013). Local significant differences from nonparametric two-sample tests. {\it Journal of Nonparametric Statistics} {\bf 25}, 635-645.

\item
Fan, J. (1996). Test of significance based on wavelet thresholding and Neyman's truncation. {\it Journal of the American Statistical Association} {\bf 91}, 674-688.

\item
Fisher, N.I. (1983). Graphical methods in nonparametric statistics: A review and annotated bibliography. {\it International Staistical Review} {\bf 51}, 25-58.

\item
Franco-Pereira, A.M., Nakas, C.T. $\&$ Pardo, M.C. (2020). Biomarker assessment in ROC curve analysis using the length of the curve as an index of diagnostic accuracy: the binormal model framework. {\it AStA Advances in Statistical Analysis} {\bf 104}, 625-647.

\item
Goldman, M. $\&$ Kaplan, D.M. (2018). Comparing distributions by multiple testing across quantiles or cdf values. {\it Journal of  Econometrics} {\bf 206}, 143–166.

\item
Inglot, T., Ledwina, T. $\&$  \'Cmiel, B. (2019). Intermediate efficiency in nonparametric testing problems with an application to some weighted statistics, {\it ESAIM: Probability and Statistics} {\bf 23}, 697-738.

\item
Janic-Wr\'oblewska, A. $\&$  Ledwina, T. (2000). Data driven rank tests for two-sample problem. {\it Scandinavian Journal of Statistics} {\bf 27}, 281-297.

\item
Kodalci, L. $\&$ Thas, O. (2024). Neutralise: An open science initiative for neutral comparison of two-sample tests. {\it Biometrical Journal} {\bf 66}, doi: 10.1002/bimj.202200237.

\item
Konstantinou, K., Mrkvi\v{c}ka, T. $\&$ Myllym\"{a}ki, M. (2024). Graphical $n$-sample tests of correspondence of distributions. {\it arXiv:2403.01838v1 [stat.ME]}

\item
Ledwina, T. $\&$  Wy{\l}upek, G. (2012a). Nonparametric tests for first order stochastic dominance. {\it Test} {\bf 21}, 730-756.

\item
Ledwina, T. $\&$  Wy{\l}upek, G. (2012b). Two-sample test for one-sided alternative. {\it Scandinavian Journal of Statistics} {\bf 39}, 358-381.

\item
Ledwina, T. $\&$  Wy{\l}upek, G. (2013). Tests for first-order stochastic dominance. {\it Preprint IM PAN} 746.

\item
Ledwina, T. $\&$ Zagda\'nski, A. (2024). ODC and ROC curves, comparison curves and stochastic dominance. {\it International Statistical Review} {\bf 92}, 431-454.

\item
Mason, D.M. $\&$ Schuenemeyer, J.H. (1983).  A modified Kolmogorov-Smirnov test sensitive to tail alternatives. {\it Annals of Statistics} {\bf 11}, 933-946.

\item
Miller, R. $\&$ Siegmund, D. (1982). Maximally selected chi square statistics. {\it Biometrics} {\bf 38}, 1011-1016.

\item
Nakas, C., Yiannoutsos, C.T., Bosch, R.J. $\&$ Moyssiadis, C. (2003). Assessment of diagnostic markers by goodness-of-fit tests. {\it Statistics in Medicine} {\bf 22}, 2503–2513.

\item
Neuhaus, G. (1987). Local asymptotics for linear rank statistics with estimated score functions. {\it Annals of Statistics} {\bf 15}, 491-512.

\item
Pardo, M. C. $\&$ Franco-Pereira, A.M. (2017). Non parametric ROC summary statistics. {\it REVSTAT} {\bf 15}, 583-600.

\item
Pettitt A.N. (1976). A two-sample Anderson-Darling rank statistic. {\it Biometrika} {\bf 63}, 161-168.

\item
Pyke, R. $\&$  Shorack, G.R. (1968). Weak convergence of a two-sample empirical process and a new approach to Chernoff-Savage theorems. {\it Annals of Mathematical Statistics} {\bf 39}, 755-771.

\item
Rousselet, G. A., Pernet, C. R. $\&$ Wilcox, R.R. (2017). Beyond differences in means: Robust graphical methods to compare two groups in neuroscience. {\it European Journal of Neuroscience} {\bf 46}, 1738-1748.

\item
Scholz, F.W. $\&$ Stephens, M.A. (1987). K-sample Anderson-Darling tests. {\it Journal of the American Statistical Association} {\bf 82}, 918-924.

\item
Song, X. $\&$ Xiao, Z. (2022). On smooth tests for the equality of distributions. {\it Econometric Theory} {\bf 38}, 194-208.

\item
Talebi, V. $\&$ Baker, C.L.Jr. (2016). Categorically distinct types of receptive fields in early visual cortex. {\it Journal of Neurophysiology} {\bf 115}, 2556-2576.

\item
Tang, L.L., Meng, Z. $\&$ Li, Q. (2021). A ROC-based test for evaluating group difference with an application to neonatal audiology screening. {\it Statistics in Medicine} {\bf 40}, 4597-4608.

\item
Zhou, W.-X., Zheng, C. $\&$ Zhang, Z. (2017). Two-sample smooth tests for the equality of distributions. {\it Bernoulli} {\bf 23}, 951-989.

\item
Wilcox, R.R. (1995). Comparing two independent groups via multiple quantiles. {\it The Statistician (Journal of the Royal Statistical Society, Ser. C)} {\bf 44}, 91-99.

\item
Wilcox, R. R. $\&$ Rousselet, G. A. (2023). An updated guide to robust statistical methods in neuroscience. {\it Current Protocols} {\bf 3}, e719. doi: 10.1002/cpz1.719

\end{description}
\end{spacing}

\end{document}